\documentclass[12pt]{aastex}
	\usepackage{epsfig}
	\usepackage{graphicx}
	\usepackage{subfigure}
	\usepackage{amssymb}  
	\usepackage{aas_macros} 
	\usepackage{amssymb}
	\usepackage{natbib}
	\usepackage{txfonts}
	\usepackage{times}
	\bibpunct{(}{)}{;}{a}{}{,} 
\shorttitle{Coronal fuzziness in multistranded loops}
\begin{document}
%
\title{Coronal fuzziness modelled with pulse-heated multistranded loop systems}
\author{Massimiliano Guarrasi\altaffilmark{1}}
\author{Fabio Reale\altaffilmark{1}}
\author{Giovanni Peres\altaffilmark{1}}
\affil{Dipartimento di Scienze Fisiche \& Astronomiche, Universit\`a di
       Palermo, Sezione di Astronomia, Piazza del Parlamento 1, 90134 Palermo,
       Italy}
\altaffiltext{1}{also INAF - Osservatorio Astronomico di Palermo ``G.S.
       Vaiana'', Piazza del Parlamento 1, 90134 Palermo, Italy}
%
\begin{abstract}
Coronal active regions are observed to get fuzzier and fuzzier (i.e. more and more confused and uniform) in harder  and harder energy bands or lines. We explain this evidence as due to the fine multi-temperature structure of
coronal loops. To this end, we model bundles of loops made of thin strands, each heated by short and intense heat pulses. For simplicity, we assume that the heat pulses are
all equal and triggered only once in each strand at a random time. The
pulse intensity and cadence are selected so as to have steady
active region loops ($\sim 3$ MK), on the average. We compute the evolution
of the confined heated plasma with a hydrodynamic loop model.
We then compute the emission along each strand in several spectral
lines, from cool ($\leq 1$ MK), to warm ($2-3$ MK) lines, detectable
with Hinode/EIS, to hot X-ray lines. The strands are then put
side-by-side to construct an active region loop bundle. We find
that in the warm lines ($2-3$ MK) the loop emission fills all the available image surface. 
Therefore the emission appears quite uniform and it is difficult
to resolve the single loops, while in the cool lines the loops
are considerably more contrasted and the region is less fuzzy. The
main reasons for this effect are that, during their evolution, i.e. pulse 
heating and slow cooling, each strand spends a
relatively long time at temperatures
around $2-3$ MK, and that it has a high emission measure during that phase, so the whole region appears more uniform or smudged. We make the prediction that the fuzziness should
be reduced in the hot UV and X-ray lines.
\end{abstract}
%
\section{Introduction}
\label{intro}
Coronal loops form the basic building blocks of the solar corona. The average temperature of these structures ranges from less than $1$ MK in cool loops to more than $10$ MK during flares \citep{Vaiana_al_1973}. The plasma inside coronal loops emits radiation mainly in the EUV and X  energy band, with spectra depending on temperature. Hence, multiband and/or multiline observations are fundamental for the investigation of the thermal structure of the corona.

Comparison of observations in different bands and different lines has shown that the confined corona, and in particular active regions, appear to be fuzzier in hotter bands and lines, better defined in cooler ones (e.g. \citealt{Brickhouse_Schmelz_2006}; \citealt{Tripathi_2009}). More specifically, according to \cite{Tripathi_2009},
a loop bundle is fuzzier when it is difficult to resolve the single loops and less fuzzy when the separations between loops are more clear and with a higher contrast. This effect has been well observed with imaging instruments.

Images taken with the 284~$\AA$~ passband filter of the Transition Region and Coronal Explorer (TRACE; \citealt{Handy_1999}), and of the EUV Image Telescope (EIT) on the Solar and Heliospheric Observatory (SOHO; \citealt{Domingo_1995}) appear to be fuzzier than in the other filters sensitive to cooler plasma.
\cite{Brickhouse_Schmelz_2006} found that resonance scattering can be ruled out as the cause of  this fuzziness.
They also suggest that it may be due to the effect of filamentation, and particularly to the high emission measure between 2 and 3 MK. Some preliminary modeling was performed for TRACE observation by \cite{Patsourakos_al_2002}.

The Extreme-ultraviolet Imaging Spectrometer (EIS; \citealt{Culhane_2007}), on board  of Hinode spacecraft \citep{Kosugi_2007}, puts this evidence under a more quantitative and objective framework \citep{Tripathi_2009}, discovering that the same active region structures are clearly discernible in cooler lines ( $\sim 1 ~ MK$), and are fuzzy at a higher temperature ($\sim 2 ~ MK$).

The question we investigate here is whether the increasing fuzziness of coronal images in hotter and hotter bands is really not an instrumental effect, but it is an intrinsic feature of the structure of coronal loop systems, and in particular to their fine transversal spatial and thermal structuring.
In this sense, we will explore the hypothesis that the same loop system looks fuzzier and fuzzier at higher and higher temperatures, and not that this effect is due to two different classes of loops, one cooler and more clearly discernible, and the other hotter and fuzzier \citep{Sakamoto_al_2009}. We will model coronal loops as composed by bundles of unresolved strands each ignited at random time by a rapid heat pulse, and compare their appearance in different spectral lines. Impulsively heated strands are commonly invoked in the framework of the nanoflare coronal heating model \citep{Parker_1988,Cargill_2004}.
The evolution of a single loop strand will be computed in detail by means of a numerical hydrodynamic loop model, and we will then form a loop system by collecting ensembles of strands. Each strand, at a given time, will be at a different phase of its evolution and will therefore emit differently. Globally the loop system will look different when observed in different spectral lines and we will study if the difference  explains the different degree of fuzziness.

In Sec. \ref{sec:model} we describe the model, in Sec. \ref{results} the results are shown and discussed in Sec. \ref{sec:disc}.

\section{The model}
\label{sec:model}

We address a typical loop system of an active region. The system is a bundle of similar loops, each consisting of many thin strands. Each strand is pulse-heated. Before the heating, the strand is tenuous and cold, actually invisible.  The loops consist of multitudes of strands, and, in principle, we should compute the hydrodynamic evolution for each of them. Instead of doing this, we make the simplifying assumption that the loops are made of similar strands and that each strand is heated by the same identical heat pulse, and therefore undergoes the same identical evolution. 
The heating rate averaged over the whole loop system is that of a steady-state loop at $3$ MK, i.e. a typical active region temperature. We assume a constant rate of heat pulses across the loops. Therefore, the steady state is reached after a transient in which the number of strands with an on-going heat pulse increases linearly. The loop half-length is $L = 3 \times 10^9$ cm. We assume that the loops stand vertical on the solar surface.

The plasma confined in each strand transports energy and moves only along the magnetic field lines, so that each strand can be described with a 1D hydrodynamic model  (e.g. \citealt{ Nagai_1980, Peres_1982, Doschek_al_1982, Nagai_Emslie_1984, Fisher_al_1985, MacNeice_1986, Gan_al_1991, Hansteen_1993, Betta_1997, Antiochos_al_1999, Muller_al_2003, Bradshaw_Mason_2003, Bradshaw_Cargill_2006}), through the equations \citep{Peres_1982,Betta_1997}:

\begin{equation}
	\frac{d n}{d t} ~ = -n \frac{\partial \textrm{v}}{\partial s}
\end{equation}

\begin{equation}
	n m_H \frac{d \textrm{v}}{d t} ~ = -\frac{\partial p}{\partial s} + n m_H g + \frac{\partial }{\partial s} \left( \mu  \frac{\partial \textrm{v}}{\partial s} \right) 
\end{equation}

\begin{equation}
	\frac{d \varepsilon}{d t} + w \frac{\partial \textrm{v}}{\partial s} ~ = Q - n^2 \beta P(T) + \mu \left( \frac{\partial \textrm{v}}{\partial s} \right)^{2} + \frac{\partial }{\partial s} \left(  \kappa T^{5/2} \frac{\partial T}{\partial s} \right) 
\end{equation}

\begin{equation}
	p ~ = \left( 1+\beta \right) n K_B T
\end{equation}

\begin{equation}
	\varepsilon ~ = \frac{3}{2} p + n \beta \chi
\end{equation}

\begin{equation}
	w ~ = \frac{5}{2} p + n \beta \chi
\end{equation}

where $n$ is the hydrogen number density; $t$ is the time, 
$s$ is the field line coordinate; $\textrm{v}$ is the plasma velocity; 
$m_H$ is the mass of hydrogen atom; $p$ is the pressure; 
$g$ is the component of gravity along the field line; 
$\mu$ is the effective coefficient of compressional viscosity (including numerical viscosity); $\beta = n_e / n$ is the ionization fraction 
where $n_e$ is the electron density; $T$ is the temperature; $\kappa$ 
is the thermal conductivity 
($\simeq 9 \cdot 10^{-7} ~ erg ~cm^{-1} ~s^{-1} ~K^{-7/2}$); 
$K_B$ is the Boltzmann constant; $\chi$ is the hydrogen 
ionization potential; $P\left( T \right) $ are the radiative 
losses per unit emission measure \citep{Raymond_1977}; 
$Q \left( s,t \right) $ is the volumetric power input 
to the solar atmosphere:

\begin{equation}
	Q \left( s,t \right) ~ = H_{steady} + H_0 f\left( t \right) g\left( s \right)  
\end{equation}

$H_{steady}$ is the steady heating term which balances radiative 
and conductive losses for the static initial atmosphere; the second term 
describes the heat pulse as a separable function of space and time.

As mentioned above, the initial atmosphere of all the  strands is tenuous and cool, since we want them to be virtually invisible in any relevant spectral band. On the other hand, the initial atmosphere has anyhow to sustain a high input energy; in particular, the chromosphere has to provide a mass amount large enough to support the evaporation driven by the heat pulse. This requires that the strand pressure cannot be too low. Apart from this, the fine details of the initial strand are not important because the energy input from the heat pulse overwhelms completely the initial energy budget and the evolution is largely independent of the initial atmosphere. A good compromise solution for this technical issue has been a loop strand with a base pressure of $0.55 \cdot 10^{-1}$ dyn $cm^{-2}$, which results in an apex temperature of $8.0 \times 10^{5} ~ K$. Since the emission of such a strand would be low but not be completely negligible in UV lines such as Mg VII and Si VI, for the sake of clarity we have decided anyway only for the analysis to set the emission of the initial strands to zero at any wavelength.
For the chromospheric part of the strand we use model F in \citet{Vernazza_1981}, appropriate for active regions. The energy balance is strictly maintained at all times also in the chromosphere.

The heat pulse is distributed uniformly along the loop strand:

\begin{equation}
  g\left( s \right) = \left\{
    \begin{array}{ll}
       1 & ~~ \mbox{ if } ~ 0 \le s \le L \\
       0 & ~~ \mbox{ elsewhere}
    \end{array} \right.
 \label{gauss_1_simul}
\end{equation}

The results do not change significantly if the heat pulses are deposited at the loop footpoints, instead of uniformly, one of the options explored by \cite{Reale_2008}.

The time dependence is a pulse function:

\begin{equation}
  f\left( t \right) = \left\{
    \begin{array}{ll}
       1 & ~~ \mbox{ if } ~ t_0 \le t \le t_0 + 60 ~  sec\\
       0 & ~~ \mbox{any other t}
    \end{array} \right.
 \label{gauss_2_simul}
\end{equation}

where  $t_0$ is the start time of heat pulse. The amplitude of the pulse is $H_0 = 0.38 ~ erg ~ cm^{-3} ~ s^{-1}$. Note that the duration of heat pulse is short enough to have a multi-temperature loop system, and not short enough to have large effects on the overall evolution and emission due to non-equilibrium of ionization \citep{Reale_2008}.

The equations are solved numerically by means of the Palermo-Harvard loop code \citep{Peres_1982,Betta_1997}. This is a well tested and highly stable code, used for both flaring \citep{Peres_Reale_1987,Betta_al_2001} and quiescent loops \citep{Reale_al_2000_II}. The Palermo-Harvard code has an adaptive grid \citep{Betta_1997}, to better describe the steep gradients along the strands and during the evolution. For the ease of post-processing, the code output results were interpolated on a fixed equispaced grid for the post-processing. The grid is made of 1024 cells along the strand.

The code output consists of temperature, density and velocity distributions along the loop strand sampled with a regular cadence during the strand evolution driven by the heat pulse. 

As mentioned above we model only one strand and replicate it so that one strand is different from the other only for the start time of the heat pulse, i.e. $t_0$. This choice minimizes the number of free parameters, and a single simulation was enough for our further analysis. The simulation computes the evolution of the strand for $2000$ s after the start of the heat pulse, and the solutions are sampled every $1$ s.

We have also assumed that all the strands are strictly out of phase, i.e. only one strand is switched on at any time. At time $t = 0$ all strands are switched off. Then the heat pulses gradually turn them on, one after the other. We have chosen to turn them on with constant cadence to have a smooth transition from no bright strands to maximum number of bright strands (Fig.~\ref{time_profile_T_med}).

At a certain time, part of the strands will still be off, in a few the heat pulse will be on, the others will be cooling after the heat pulse has ended. Since all strands have the same evolution, at each time each strand will be at a different phase of the same evolution. A whole loop at a given time is therefore simulated as a set of otherwise identically evolving strands, but each with a different heating start, all glued together. This approach is not completely new; outputs of the same loop simulation have been put together in the past to simulate multistrand nanoflaring loops
\citep{Peres_al_1993_inproc,Warren_al_2002,Warren_al_2003,Winebarger_al_2003_a,Winebarger_al_2003_b} but in all cases they were {\it averaged} to obtain the effective aspect and evolution of a loop as a whole. The novelty of our approach is that we do not model each loop as whole, but we keep the information of its spatial longitudinal and  transversal distribution.

In our model one heat pulse is switched on every one second in one new strand. 
One loop consists of $\approx 60$ strands and we put $\approx 30$ loops one by the other. Every second there are $60$ strands where the heat pulse is on out of a total of $2000$. In the steady-state the loop heating per unit volume is therefore $\overline{H} = {H_0} \times 60/2000 \approx 0.011$ erg ~ cm$^{-3}$ s$^{-1}$. 
According to  scaling law \citep{Rosner_1978}, this steady heating sustains a loop with an apex temperature of $ \approx 3.8 ~ MK$  corresponding to an average temperature along the loop of about $3 ~ MK$(Fig.~\ref{time_profile_T_med}).

When we put all strands side by side, we obtain a mesh $1024 \times 2000$ grid cells. The transversal width of each strand is constant along the loops. Therefore it is only a multiplicative constant which determines the cross-section of the entire loop system.

From model results we synthesized the loop emission in different spectral lines. We selected the spectral lines, in Tab. \ref{lines_used_sigma}, with a peak emissivity in a wide range of temperatures. A subset (the first seven) are  detected with Hinode/EIS (\cite{Tripathi_2009}), which provide our reference observational evidence. We also include hotter lines detectable in the UV (e.g. Fe XIX, Fe XX, Fe XXIII from AIA on board the SDO mission) and X-ray band.

\begin{figure}[htbp]               
 \centering
\subfigure
   {\includegraphics[width=13.8cm]{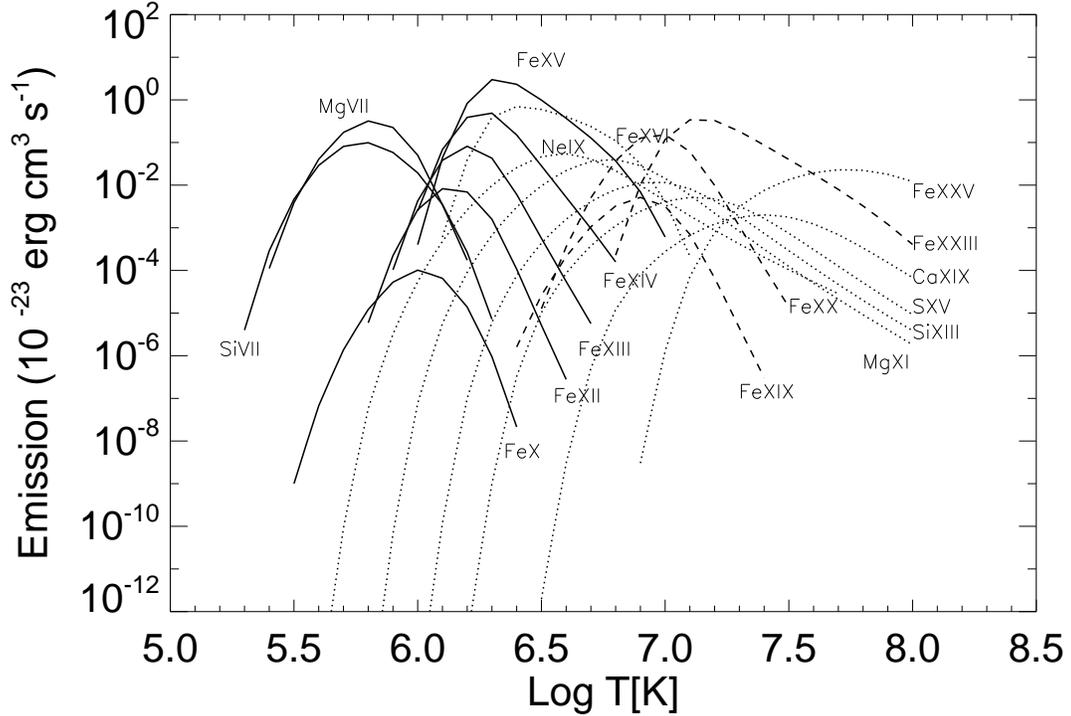}}
\caption{Profile of emission per unit of emission measure vs $log ~ T[K]$ for the various spectral lines. We make a distinction between lines for comparison with \citet{Tripathi_2009} ({\it solid lines}),  lines in AIA filter bands on board of SDO mission ({\it dashed lines}), and  hot X ray lines ({\it dotted lines}) .}
\label{lines_profile}
\end{figure}

The emissivity in these lines ($G(T)$, Fig.~\ref{lines_profile}) is taken from the CHIANTI database \citep{Chianti_I,Chianti_VI}, with coronal abundances of \cite{Feldman_1992} (assuming a density of $10^9 ~ cm^{-3}$), and ionization fraction of \cite{Mazzotta_al_1998},  .

The emission from each grid cell is computed as:

\begin{equation}
  I\left( T, n_e \right) ~ = ~ \int_{V_{pixel}} n_{e}^{2} G\left( T \right) dV
 \label{emission_formula}
\end{equation}

where $V$ is the volume and $V_{pixel}$ is the volume enclosed by a single pixel.

\begin{figure}[htbp]               
 \centering
\subfigure
    {\includegraphics[width=6.3cm]{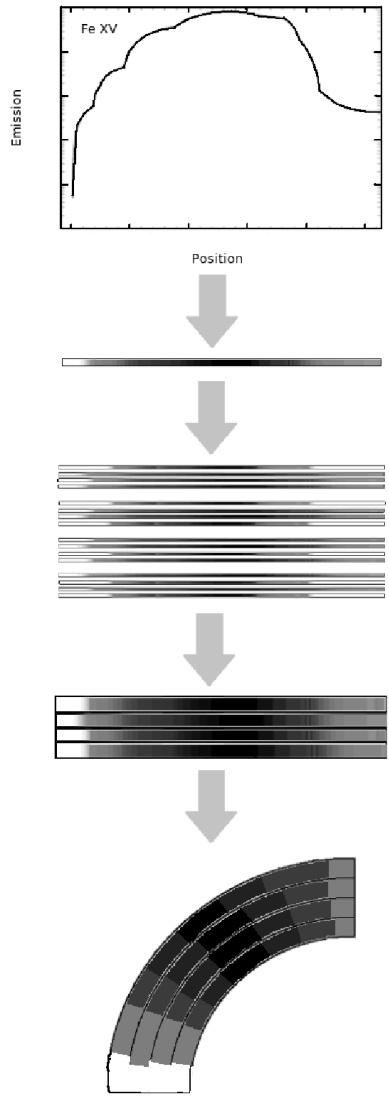}}
\caption{\small  Flowchart showing how an loop bundle image is obtained from the emission of each individual loop in a given spectral line. We first compute the spatial distribution of the emission along each strand, we then put the emission in the form of an image made of a pixel strip, we put all pixel strips side by side; we then reconstruct the aspect of the whole loop system, then we group the strands into loops. Finally we bend the image to obtain a loop-like shape.}
\label{Schema}
\end{figure}

To obtain a final image to be compared with observational data we follow the conceptual steps shown in Fig.~\ref{Schema}. We first compute the emission along each strand in a given line. The strand emissions can be put in the form of a 1-D image consisting of a strip of pixels on an appropriate grey scale. We obtain 2000 pixel strips. We then put them side by side and obtain a $1024 \times 2000$ pixel image. We have chosen to rebin our images by summing over bundles of $65$ strands, so to obtain a collection of about $30$ parallel loops.

\section{The results}
\label{results}

\begin{figure}[htbp]
 \centering
 \subfigure[]
   {\includegraphics[width=13.8cm]{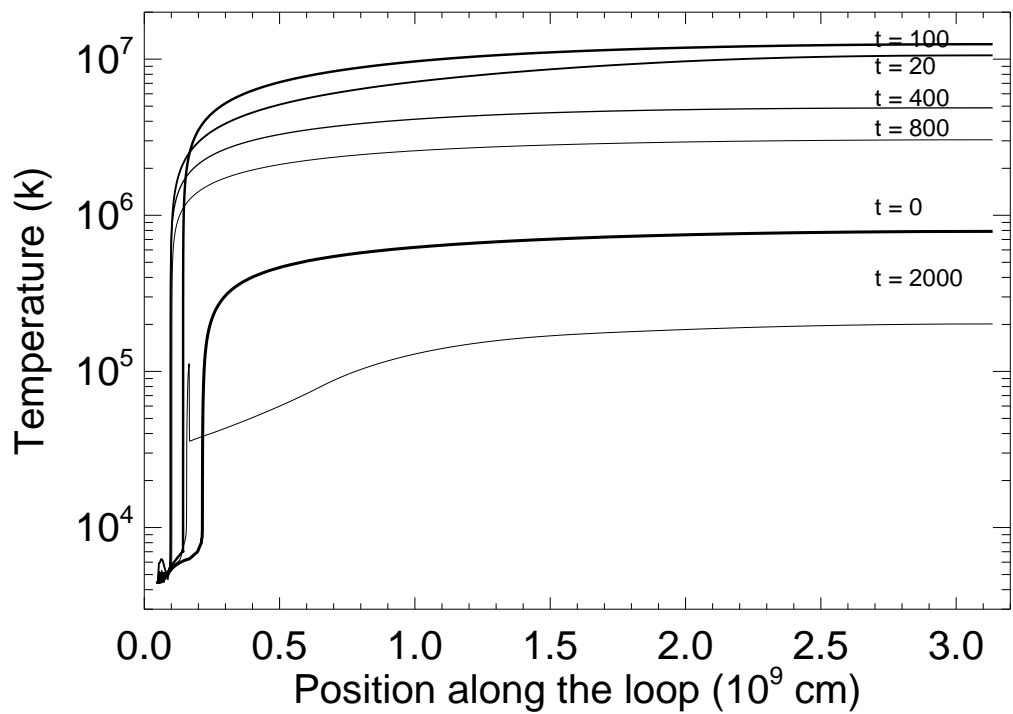}}
 \hspace{7mm}
\subfigure[]
   {\includegraphics[width=13.8cm]{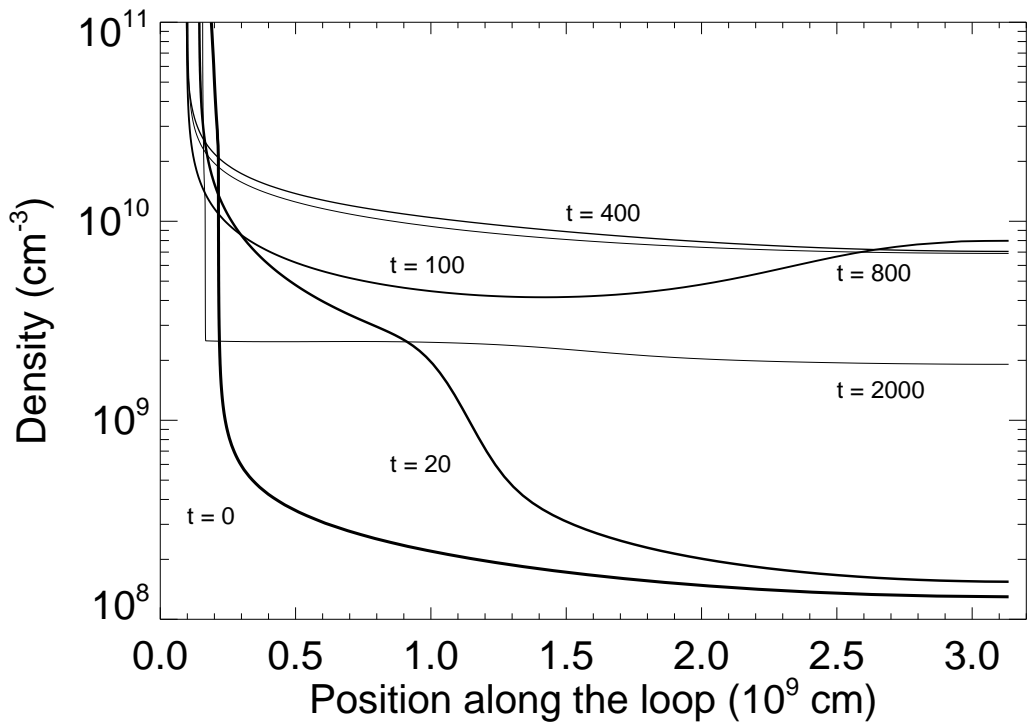}}
\caption{Temperature (a), and density (b) along half of a single strand at times $t=0$, $20$, $100$, $400$, $800$, $2000$ ~ s (thinner and thinner lines with progressing time). The heat pulse starts at time $t=0$ and lasts $60 ~ s$. }
\label{time_profile}
\end{figure}

The evolution of nanoflaring plasma confined in coronal loop is well known from previous works \citep{Peres_al_1993_inproc,Warren_al_2002,Warren_al_2003,Patsourakos_Klimchuk_2005,Testa_al_2005} and is similar to that of proper flaring loops, although on a smaller scale (e.g., \cite{Nagai_1980,Peres_1982,Reale_Peres_1995}).
We show the evolution of the density and temperature along a single strand in Fig.~\ref{time_profile}. The temperature soon settles to about 10 MK along most of the strand, due to the strong heat pulse; then it slowly decreases as expected with an e-folding time scale given by $\tau_s \approx 500 ~ L_9/\sqrt{T_6} \approx 500$ s (Reale 2007, 2009) where $T_6$ and $L_9$ are the maximum temperature and the length of the strand measured in units of $10^6$ K (MK) and $10^9$ cm respectively. At $t = 2000$ s the strand has cooled below the temperature it had before the heat pulse.  At $t = 20$ s the density plot clearly shows a strong evaporation front coming up from the chromosphere (the density of the front jumps by more than a factor 10). At $t= 100$ s the front has filled the whole strand with some extra accumulation in the apex region. At $t = 400$ s the density distribution has a shape similar to the initial equilibrium one, but settled around a maximum value of $\sim 10^{10}$ cm$^{-3}$. Then the density begins to decrease and at the final time computed it is lower, by a factor $\sim 5$ than the maximum, still much higher than the initial value.

\begin{figure}[htbp]
 \centering
\subfigure
   {\includegraphics[width=13.8cm]{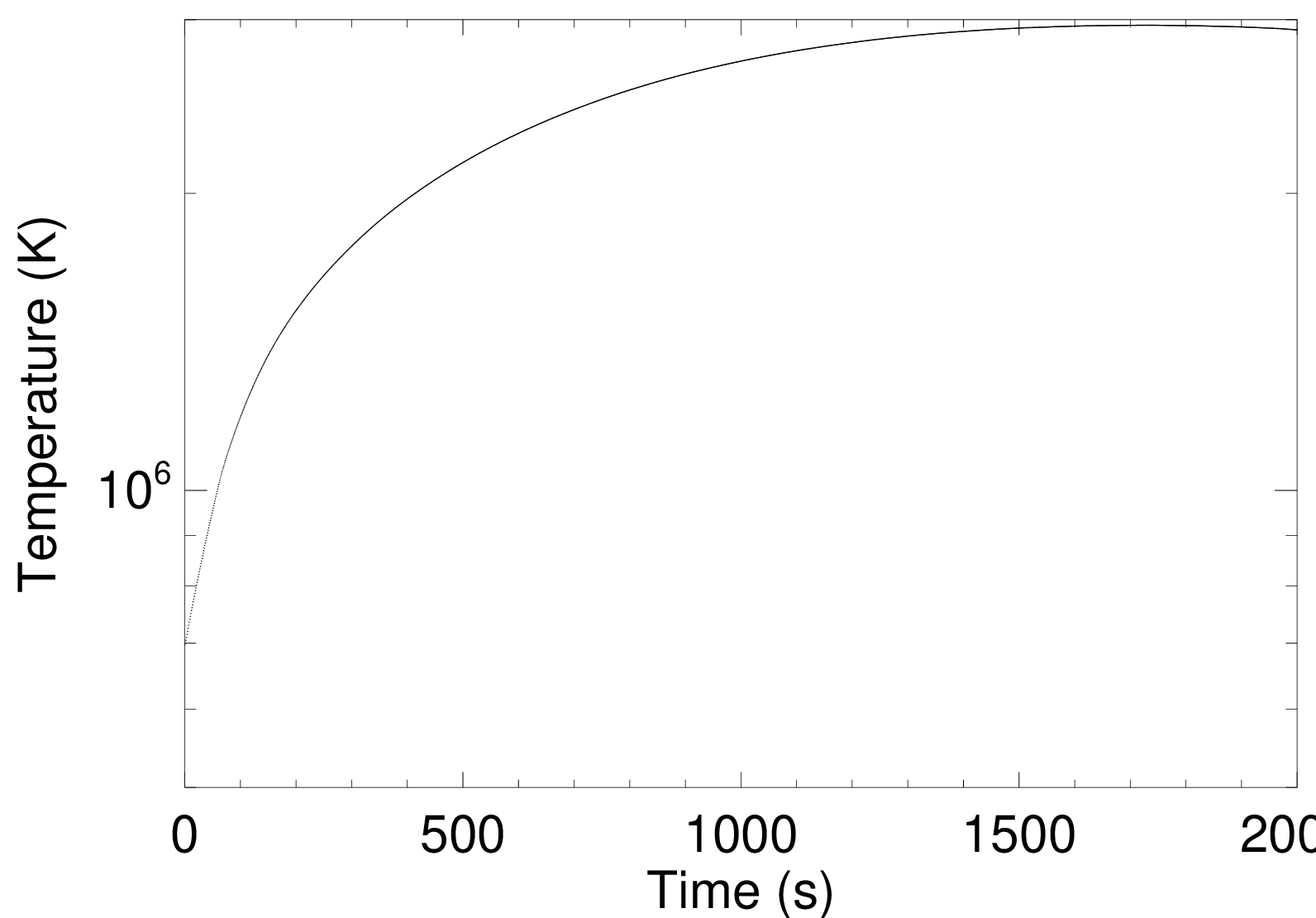}}
\caption{Evolution of the loop average temperature from off-state to steady-state.}
\label{time_profile_T_med}
\end{figure}

As described in Sec.~\ref{sec:model}, this evolution is replicated in all strands with different relative timing, depending on the time the heat pulse is switched on. As mentioned above, for a realistic situation, we have simulated the ignition of the whole loop system, with a gradually increasing number of simultaneously heated strands. The transient evolution will be the subject of a future work, but Fig.~\ref{time_profile_T_med}  shows the initial evolution of the temperature averaged over the whole loop system. In the following we will focus our attention to the final time ($t= 2000$ s) when the loop system enters a presumably long steady state, in which the heat pulse 'storm' (i.e. the 2000 pulses in total) repeats continuously. In a way, we are therefore implicitly assuming that the time taken to re-energise the magnetic field (e.g. by twisting, braiding) is about 2000 s.

\begin{figure}[htbp]
 \centering
\subfigure
   {\includegraphics[width=13.8cm]{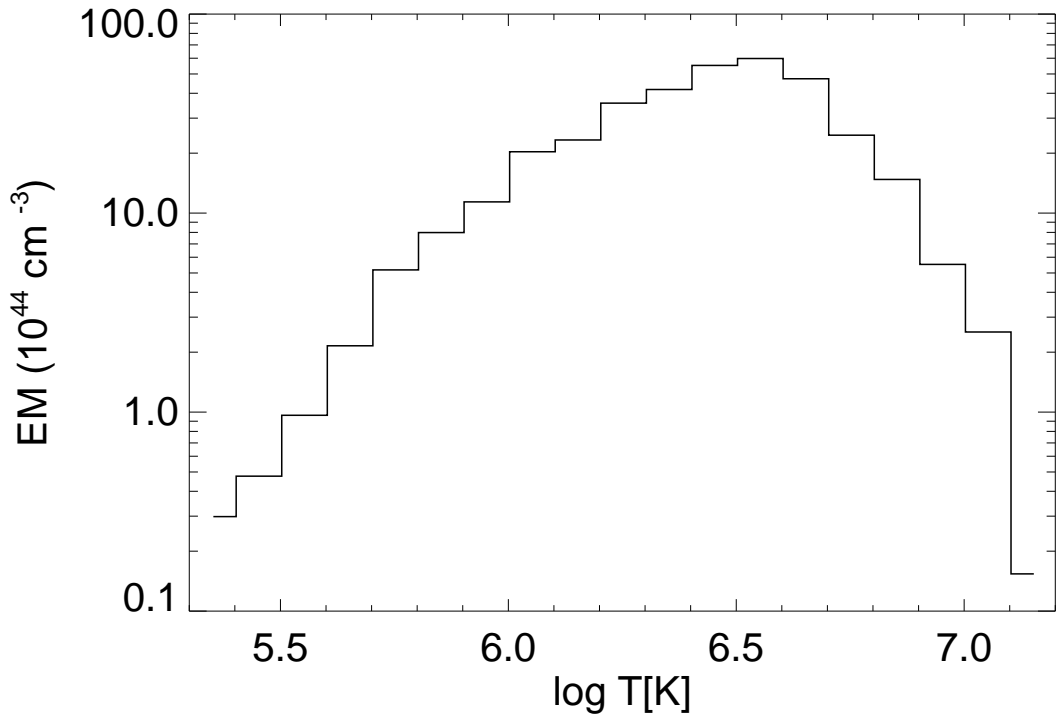}}
\caption{Distribution of emission measure vs Temperature at $t ~ = ~ 2000$ s. We show the distribution of the coronal part only, i.e. the upper 90\% of the loops.}
\label{EM_T_2000s}
\end{figure}

Fig.~\ref{EM_T_2000s} shows the distribution of the emission measure of the entire collection of strands, i.e. of the whole loop system, versus temperature (e.g. \citealt{Cargill_1994}) at this time. We show the distribution of the coronal part only, i.e. upper 90\% of the loop bundle, excluding the lower layers. The peak of the distribution is around 3 MK, as planned. We also see that the distribution is quite broad; the flatter tail is toward the cool part, but there are also significant hot components up to 10 MK, as expected, due to the presence of the strong heat pulses. These hot components are of course minor, due to the small duty cycle of the heating phase with respect to the whole evolution of the strand.

\begin{figure*}
\centering
\subfigure[{\bf Mg VII}]{\includegraphics[width=4.0cm]{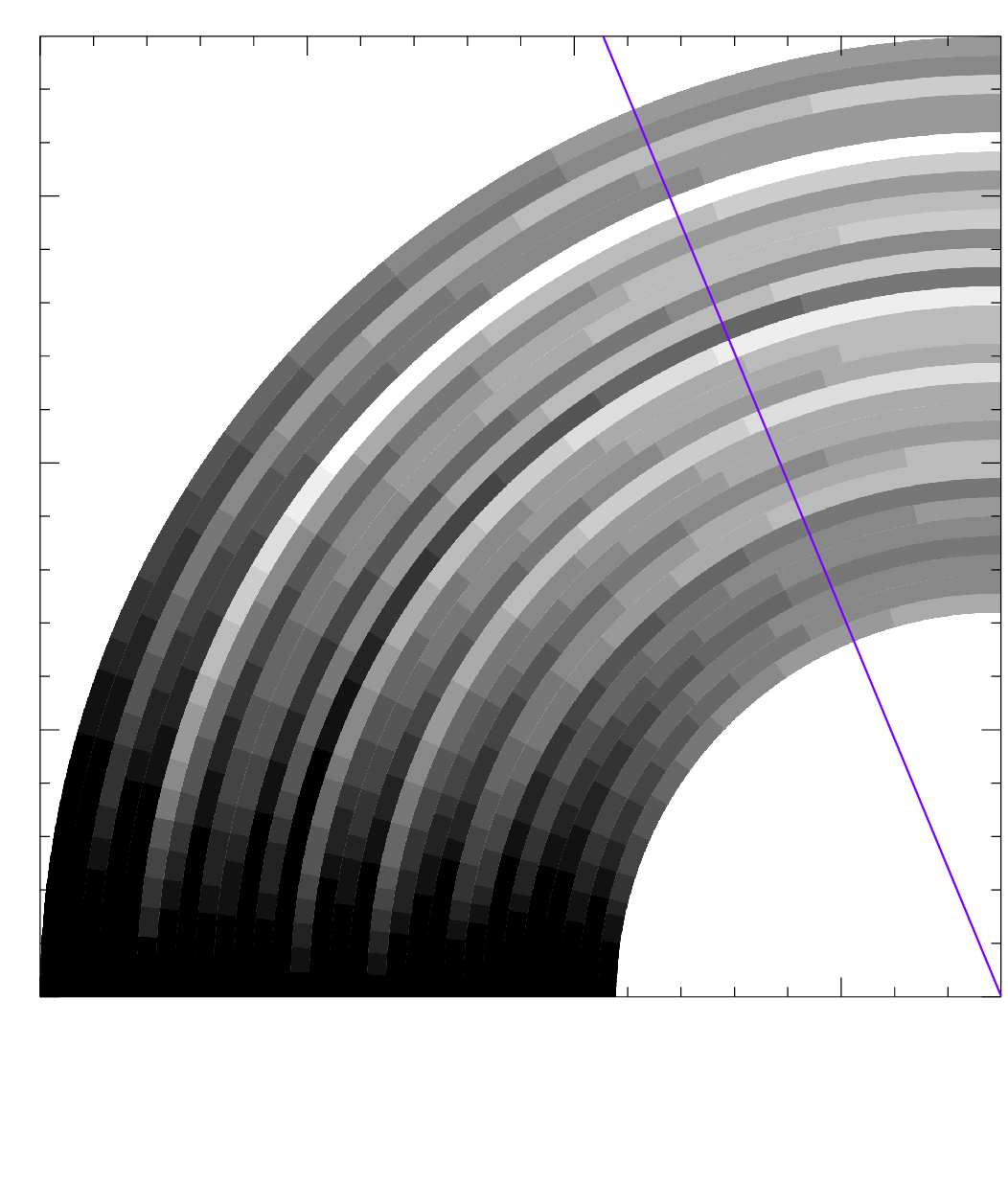}}
\hspace{2cm}
\subfigure[{\bf Mg VII}]{\includegraphics[width=5.3cm]{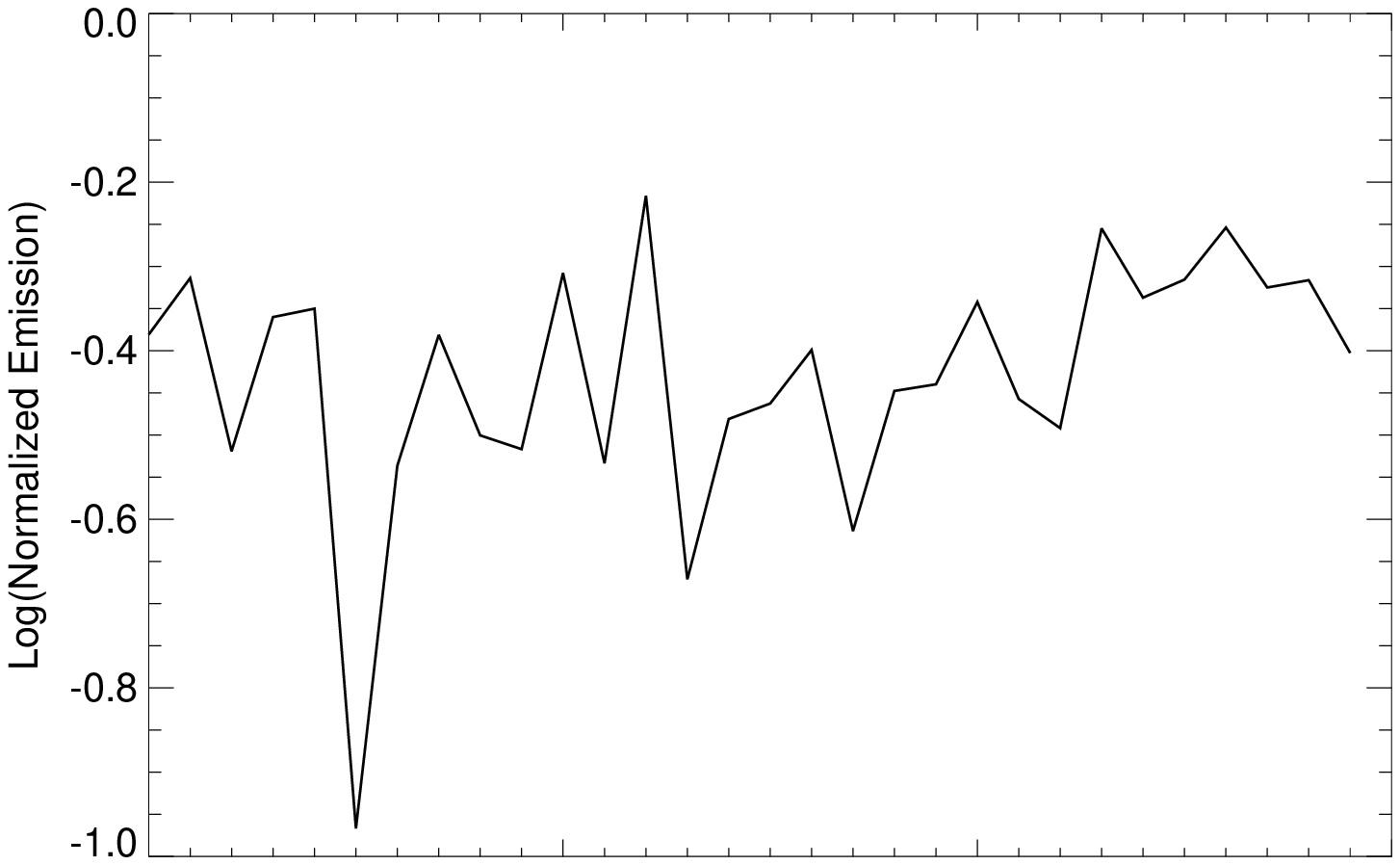}}
 \centerline{~}
\subfigure[{\bf Fe X}]{\includegraphics[width=4.0cm]{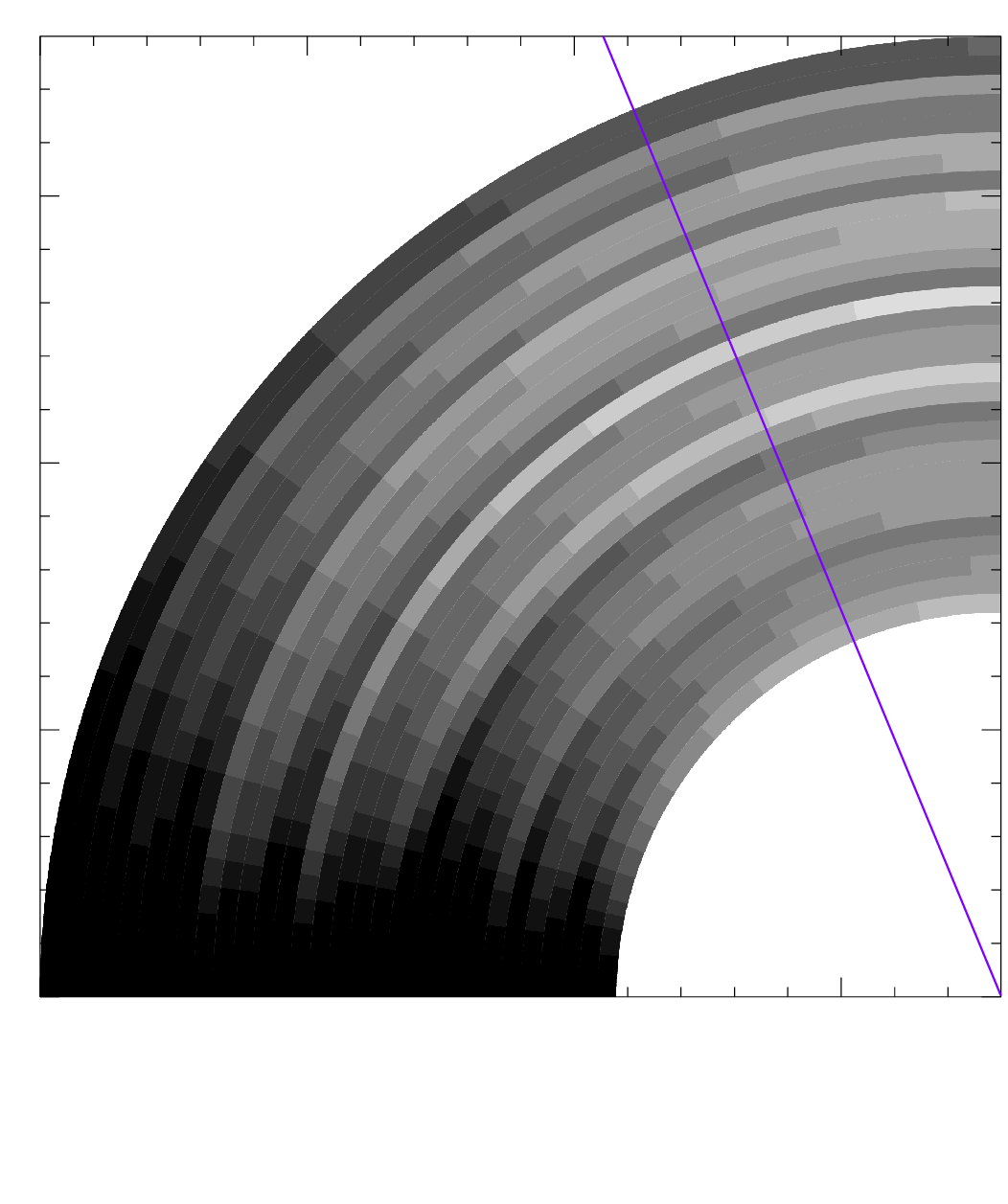}}
\hspace{2cm}
\subfigure[{\bf Fe X}]{\includegraphics[width=5.3cm]{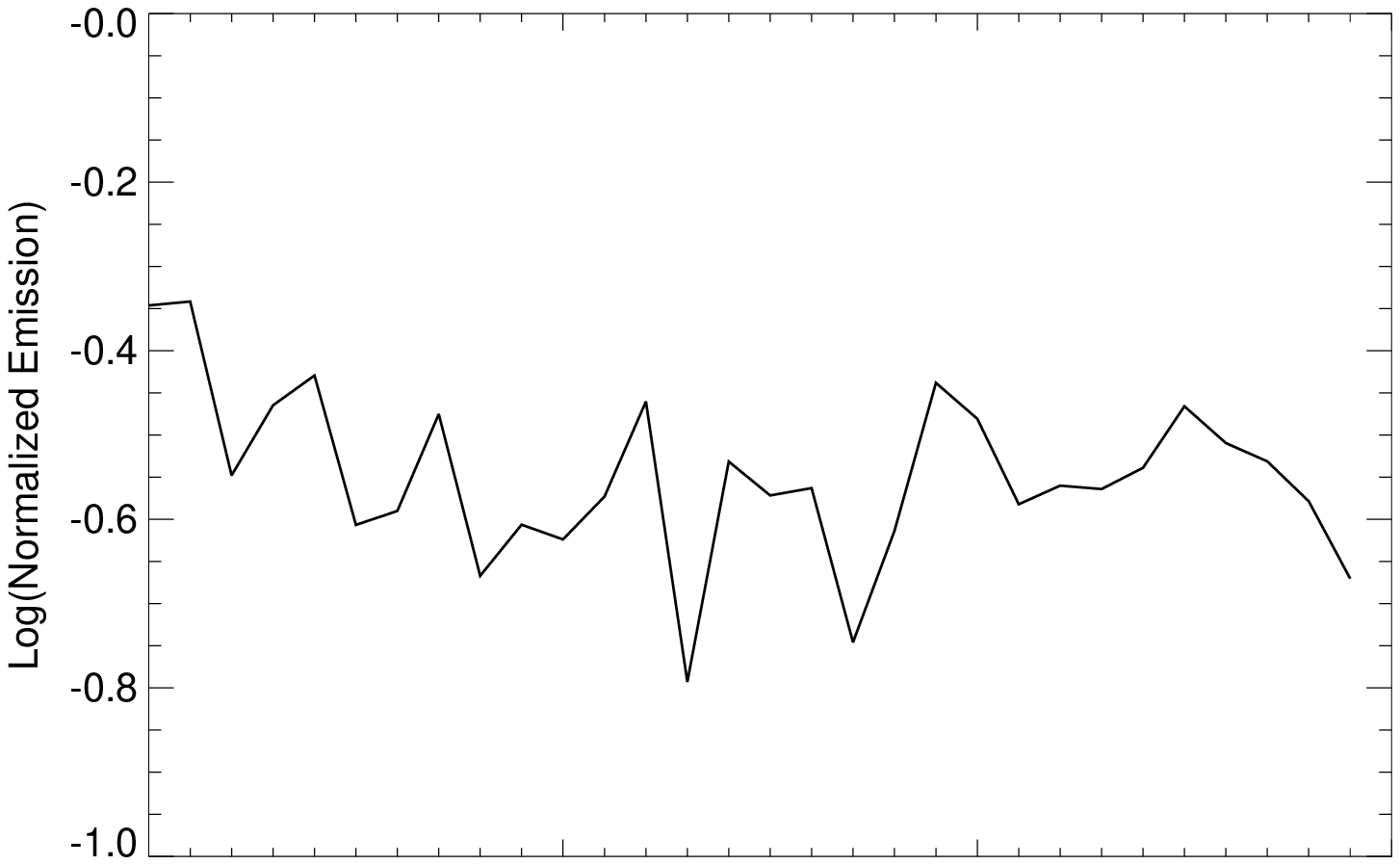}}
 \centerline{~}
\subfigure[{\bf Fe XV}]{\includegraphics[width=4.0cm]{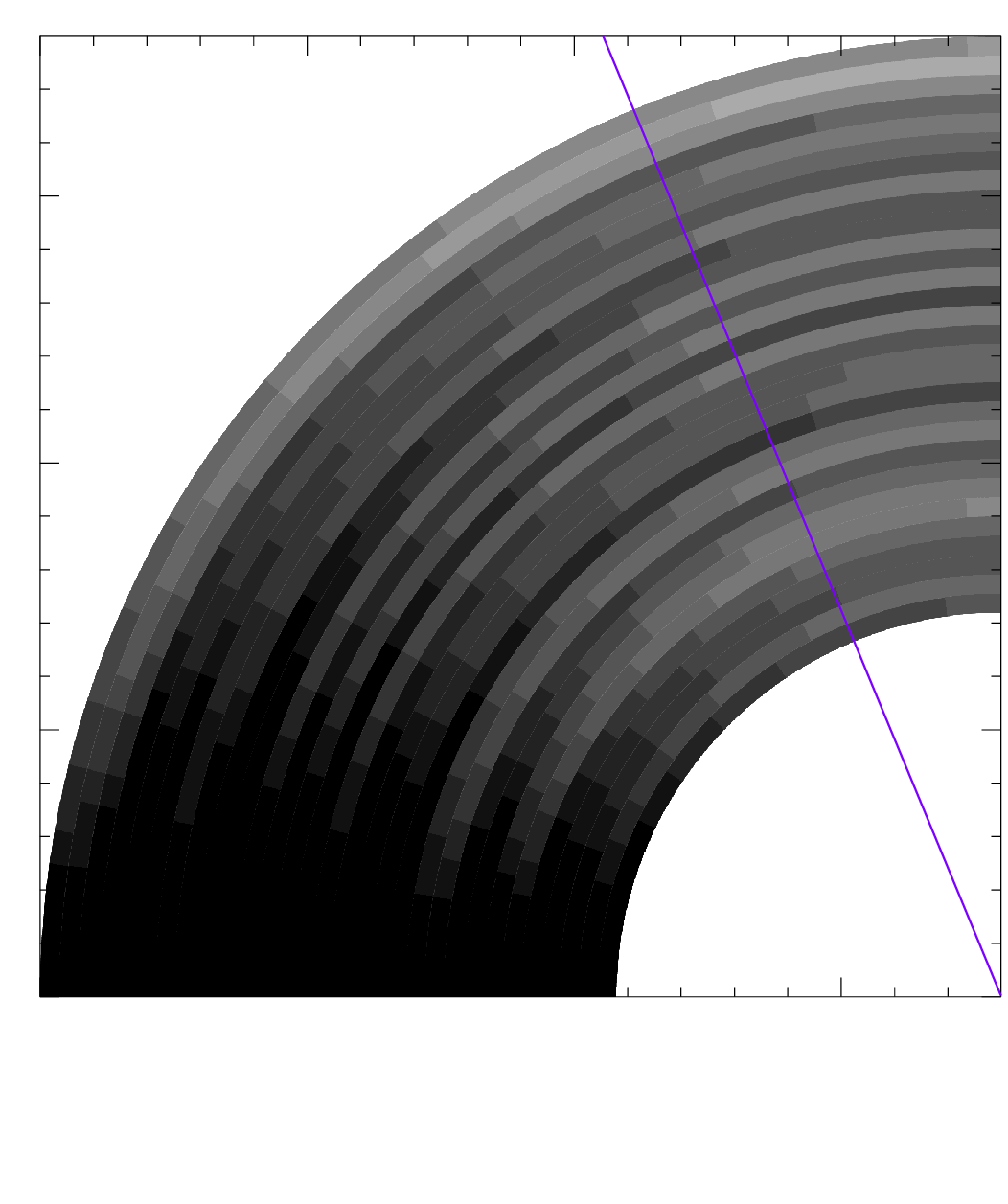}}
\hspace{2cm}
\subfigure[{\bf Fe XV}]{\includegraphics[width=5.3cm]{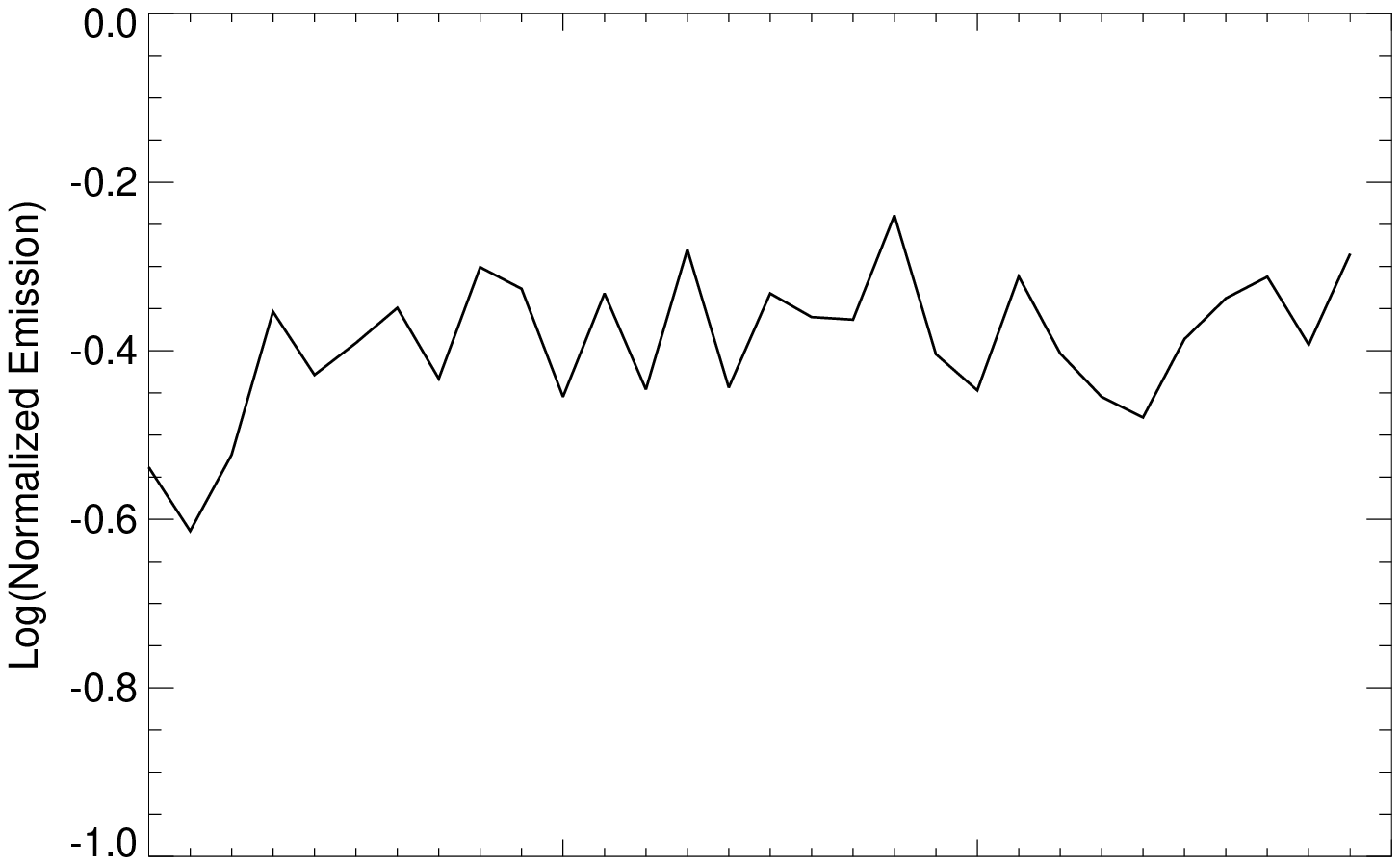}}
 \centerline{~}
\subfigure[{\bf Fe XXIII}]{\includegraphics[width=4.0cm]{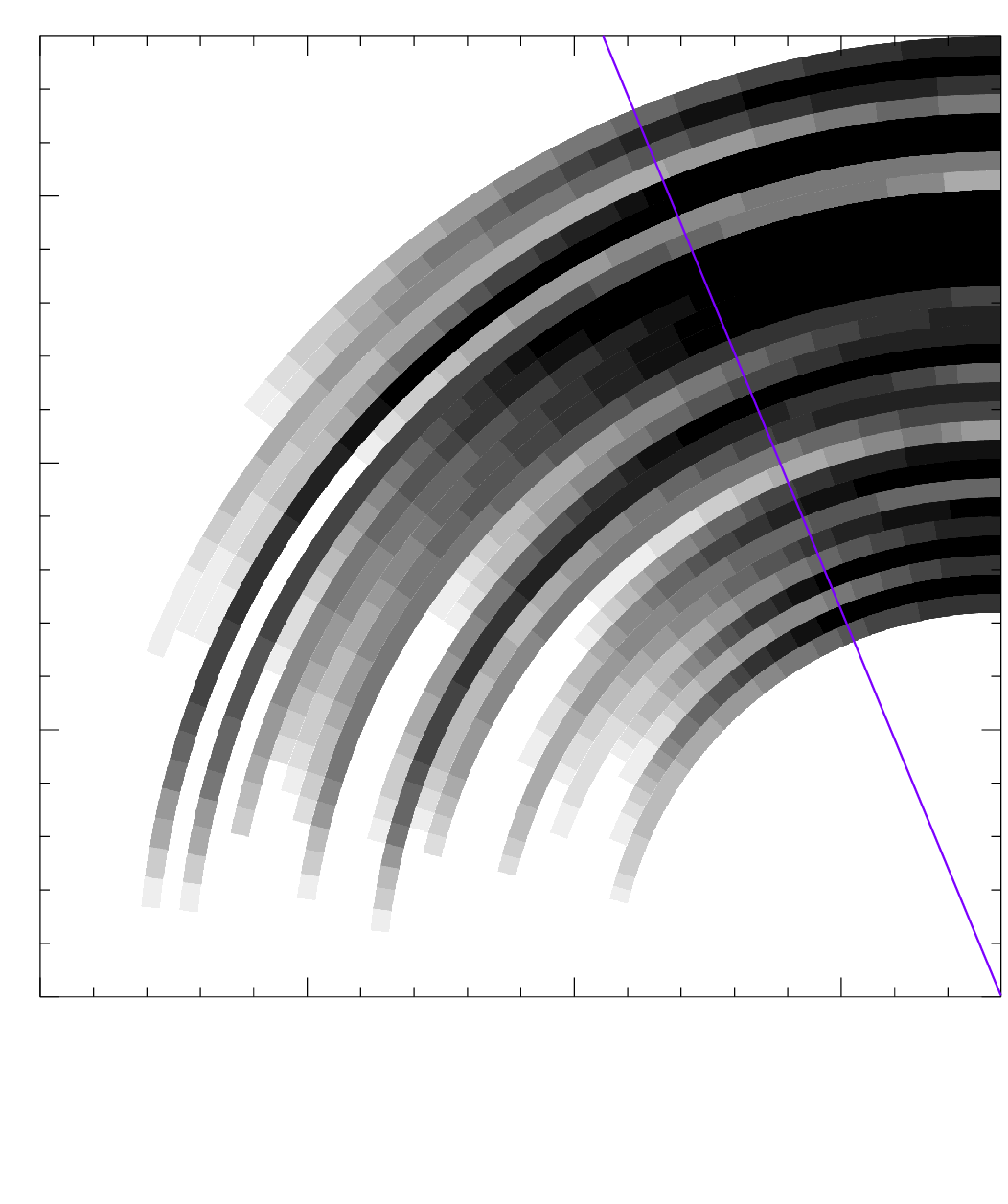}}
\hspace{2cm}
\subfigure[{\bf Fe XXIII}]{\includegraphics[width=5.3cm]{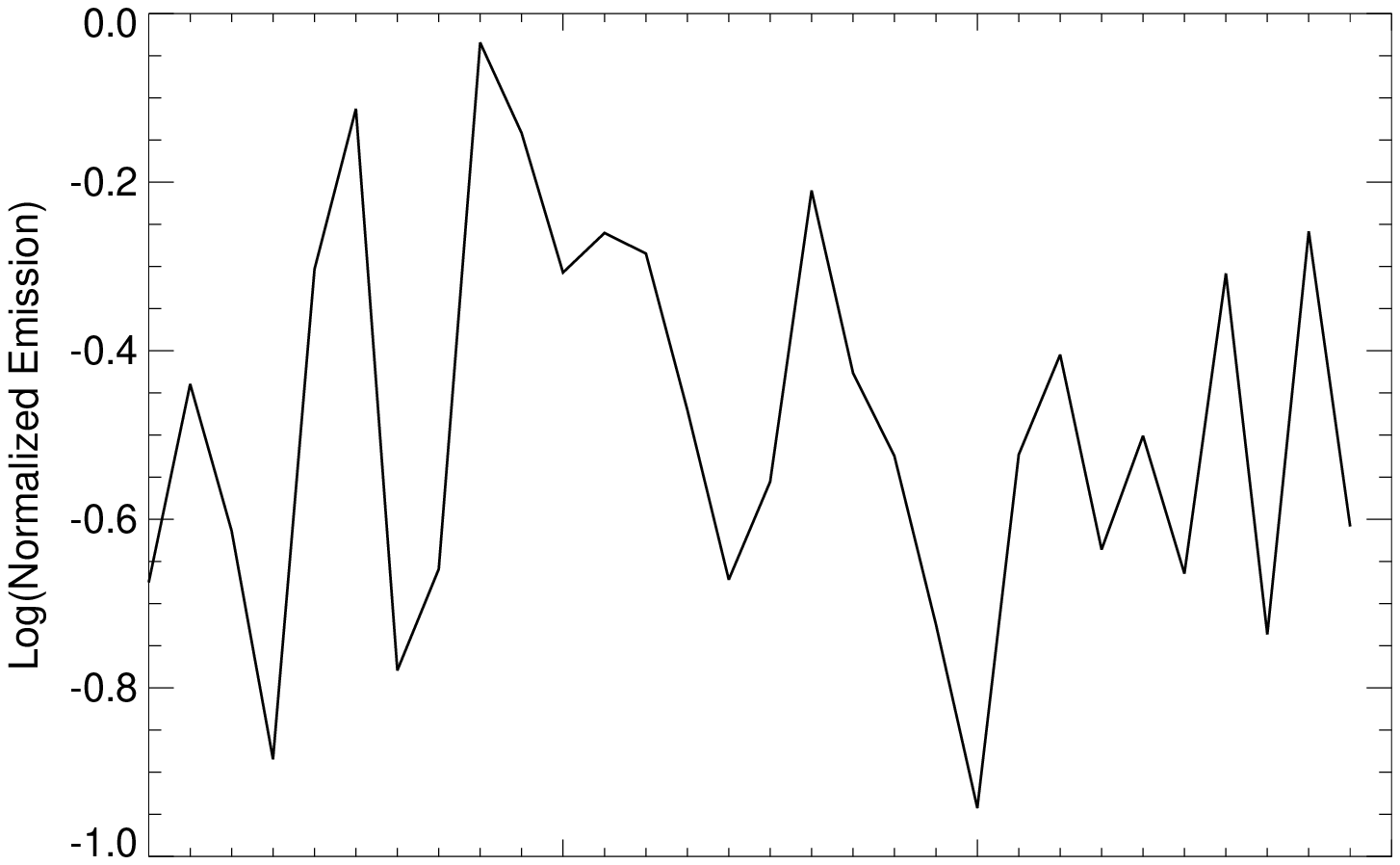}}
\caption{Loop bundle emission in different (labeled) spectral lines. In the images (left column) we show the upper $90\% $ of the loop. The grey scale is logarithmic, and covers a factor 10 of intensity; black corresponds to the maximum of emission and white to the minimum. The plots in the right column are the emission profiles along the lines marked in the images of the left column, in the same logarithmic scale. }
\label{loop_profile}
\end{figure*}

Using the method described in the Sec~\ref{sec:model}, we synthesize maps of emission for all the lines considered. Fig. \ref{loop_profile} (left column) shows a subset of them, namely the map in cool EIS lines, i.e. Mg VII $\lambda 278$ ($log ~ T[K] ~ = ~ 5.8$), medium temperature EIS line Fe X $\lambda 186$ ($log ~ T[K] ~ = ~ 6.0$), warm EIS line Fe XV $\lambda 284$ ($log ~ T[K] ~ = ~ 6.4$), and hot X-ray line Ca XIX $\lambda 3.21$ ($log ~ T[K] ~ = ~ 7.4$). The images on the left column are analogous, and can be compared, to observed ones (e.g. \cite{Tripathi_2009}). The grey scale is logarithmic and spans a factor $10$ in all maps and plots. This is a customary choice for showing observed data (e.g. \cite{Tripathi_2009}). In the first three lines, the emission decreases from the base to the top of the loop, due to the density decreasing as well. We show the intensity of the upper $90\%$ of the loop system, i.e. the coronal part only. In this way we exclude the \emph{moss} emission from cromosphere and transition region, which, nevertheless, would appear saturated in this color scale. This may indicate that the related emission measure predicted by the model is too high, which is a well known effect of 1-D hydrodynamic loop modeling (e.g. \citealt{Warren_Winebarger_2006,Warren_Winebarger_2007}), and may point to the need for a more accurate description of the low loop atmosphere. This is however not required in this context.
In the hot Fe XXIII line the opposite occurs, i.e. the brightness increases from the base to the top of the loop system, because the line is sensitive to plasma with $\log T \geq 6.9$, which is practically never found in the low part of the loop system. For our specific question, from the comparison of the first three lines we can immediately see that we can easily resolve the single loops in the cool line image, that this is more difficult in the second image, and that the Fe XV image is much more uniform and does not practically allow us to resolve the structures. Interestingly, the fourth image in the hot UV line shows again very well contrasted structures. This occurs also for the other hot lines, although with different levels of contrast. The plots in the right columns of Fig. \ref{loop_profile} quantify better this effect: the profile along the cut in the cool line shows large fluctuations, with large amplitude and high spatial frequency, in the warm line profile the fluctuations are of smaller amplitude.   According to the operative definition reported in \citet{Tripathi_2009}, the loop system in the Fe XV image is fuzzy, the loop system in the Mg VII image is not. This is in total agreement with the observational evidence shown in \citet{Tripathi_2009}.

\begin{figure}
 \centering
 \subfigure
   {\includegraphics[width=13.8cm]{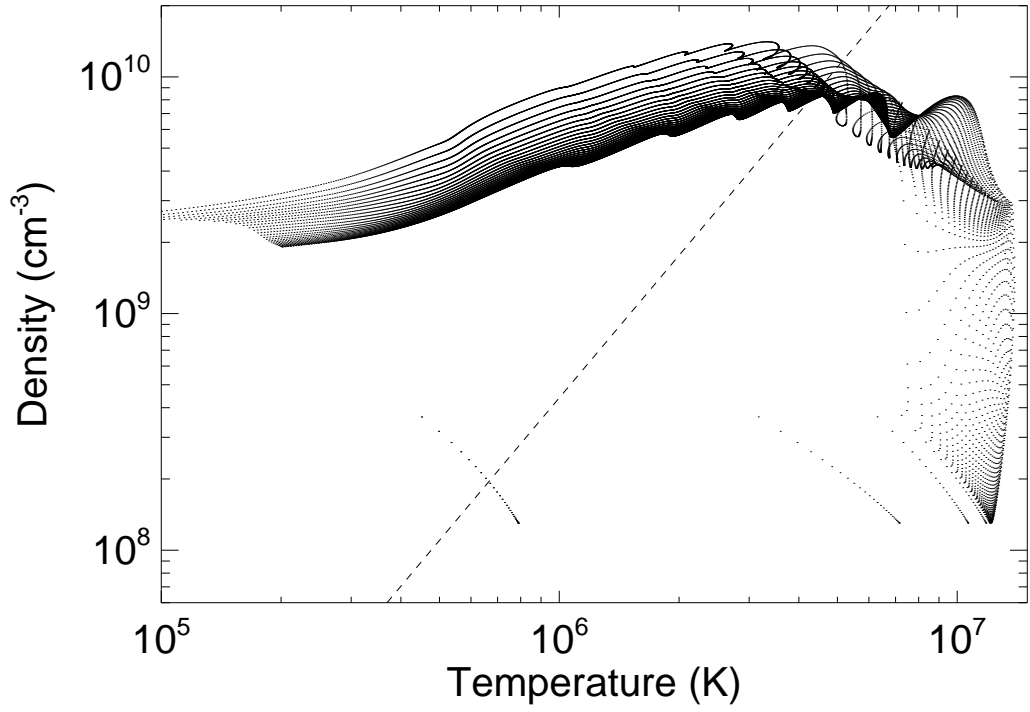}}
\caption{Scatter plot of the values of density and temperature, obtained from averaging on bins of 32 grid points in the whole model strand evolution (upper 90\% part of the loops). The dashed line is the locus of density and temperature at the apex of hydrostatic loops, with half-length $L=3.0 \times 10^{9}~$cm, according to the scaling laws of \cite{Rosner_1978}.}
\label{scatter_plot}
\end{figure}

We have also devised a figure for fuzziness to be computed in the cross-section line:

\begin{equation}
 \frac{\sigma_I}{\overline{I}} ~ = ~ \frac{1}{\overline{I}} \cdot  \sqrt{\frac{1}{n}\sum_{i=1}^{n} \left( I_i - \overline{I} \right)^{2} }
 \label{sigma_med}
\end{equation}

where $I_i$ is the emission in the spectral line in the i-th pixel along the cross-section, $\overline{I}$ is the average value for $I_i$, and $n$ is the number of pixel in the region marked by the line (left panels in Fig. \ref{loop_profile}).
Clearly, the larger this quantity, the more contrasted and well resolved are the loops, and the smaller is the fuzziness. Table \ref{lines_used_sigma} includes the values of this figure for all the spectral lines analyzed. All these value are calculated along the same cut shown in Fig. \ref{loop_profile}.

\begin{table*}[]
  \begin{minipage}[b][90mm]{\textwidth}
   \vspace{2.5truecm}
   \tabcolsep 0.25truecm
   \begin{center}
    \caption{Variation of emission in the lines analyzed.}
    \begin{tabular}{cccc}
     \\
     \hline
     \hline
     \\
     Line & Wavelength ($\AA$)&  Emission temperature ($log T[K]$) & $\sigma_I / \overline{I}$\\
     \\
     \hline
     \\
     Mg VII   & $278$  & $5.8$ & 0.28\\
     Si VII   & $275$  & $5.8$ & 0.26\\
     Fe X     & $186$  & $6.0$ & 0.23\\
     Fe XII   & $195$  & $6.1$ & 0.22\\
     Fe XIII  & $196$  & $6.2$ & 0.21\\
     Fe XIV   & $264$  & $6.3$ & 0.20\\
     Fe XV    & $284$  & $6.4$ & 0.18\\
     Fe XVI   & $361$  & $6.5$ & 0.15\\
     Ne IX    & $13.4$ & $6.6$ & 0.12\\
     Mg XI    & $9.31$ & $6.8$ & 0.14\\
     Fe XIX   & $133$  & $6.9$ & 0.26\\   
     Fe XX    & $133$  & $7.0$ & 0.39\\
     Si XIII  & $6.69$ & $7.0$ & 0.21\\
     Fe XXIII & $133$  & $7.1$ & 0.55\\
     S XV     & $5.10$ & $7.2$ & 0.30\\
     Ca XIX   & $3.21$ &$7.4$ & 0.46\\
     Fe XXV   & $1.87$ &$7.6$ & 1.0\\
     \\
     \hline
     \hline
     \label{lines_used_sigma}
    \end{tabular}
   \end{center}
  \end{minipage}
 \end{table*}

The values are quite high for the lines at low temperatures ($log T \le 6$; Mg VII, Si VII, Fe X), very high for the lines at very high temperature ($log T \ge 6.9$; e.g. S XV, Ca XIX, Fe XXIII, Fe XXV), lower for the warm lines (around $3 ~ MK$;  especially FeXV, FeXVI, Ne IX). We have applied the same analysis for different binnings, and still find similar results.

The reason why the loop system appears more uniform at warm temperatures is clear from inspection of Fig. \ref{scatter_plot}. In the figure each data point marks a value of density and temperature obtained from averaging over sections of 32 grid points along a strand (upper 90\%) in each output. The figure clearly shows that around 3 MK the density of data points is higher, and at higher values of density. The implication is that each strand spends a long time, and with a high emission measure, at an average temperature of about 3 MK, and therefore at this temperature more strands appear on the average brighter, i.e. the loop system is more uniformly bright. The plasma does not stay long at a higher temperature, i.e. around the heat pulse, and moreover during this phase, the emission measure is not  high because the evaporation is still in an early phase. The plasma does instead stay for a long time at lower temperature, but the emission measure becomes very low, not contributing much to the overall emission. Fig. \ref{scatter_plot} also shows that this model fits well the observational constraint of alternating cool/overdense - hot/underdense status of coronal loops (e.g. \citealt{Klimchuk_2006}). 

It is easy to imagine extrapolating this result to whole active regions, threaded with many thousands upon thousands of magnetic field lines. There are fewer field lines populated with low and high temperature plasma and so distinct loop structures appear. There are many more field lines populated with plasma in the 2 - 3 MK range and so fuzzy emission appearing to be composed of unresolved structures would arise as observed.

\section{Discussion and Conclusions}
\label{sec:disc}

In this work we investigate the reason why coronal loop systems, or entire active regions, look fuzzier in warm $\sim 2-3$ MK spectral lines than in cooler $\leq 1$ MK lines. This evidence has been well known since the first X-ray and UV missions, e.g. Skylab, but it has been recently put on a more established ground from Hinode/EIS spectral observation \citep{Tripathi_2009}. Our basic scenario is that coronal loops consist of bundles of thin strands, each of thickness below the instrumental spatial resolution, and that each strand is heated up to about 10 MK by a strong and fast heat pulse, i.e. the loops are heated by a storm of nanoflares. The plasma is confined in each strand, so that it evolves as an independent atmosphere, and can be modeled with loop hydrodynamics (see also \cite{Patsourakos_Klimchuk_2007} for a conceptually similar approach). Our choice has been to assume that the strands are all heated once and by the same heat pulse, lasting 60 s, occurring at a different random time for each strand, with a cadence and an intensity adequate to maintain the loop at $\sim 3$ MK on average. This model fits well the observational constraint of hot/underdense-cool/overdense cycles.  We have then collected $2000$ different strands to form a loop system, and derived synthetic images of the loop system when it reaches steady state in several relevant spectral lines. In our opinion, the images synthesized from our model unequivocally show the same ``fuzziness'' in the same warm lines as observed with EIS, and the same better definition in the cool lines as observed with EIS. In other words, our model is able to explain the evidence. Of course, it explains also the effect as observed in narrow-band XUV instruments such as the normal-incidence imaging telescopes, TRACE and SoHO/EIT. We have also provided quantitative figures to this effect. The basic reason why this model works is that, in spite of the short heat pulse, the strands spend a long time with a high emission measure at a temperature around 3 MK, much less time when plasma is hotter and long time, but with much less emission measure when the plasma is cooler.

So the loop systems appear more uniform around 3 MK, and this higher filling factor gives the impression of ``fuzziness", as described in \citet{Tripathi_2009}. In cooler lines we are able to resolve better the loops, which appear more contrasted and with better defined boundaries.

This ``fuzziness'' is different from the one intended in \citet{Sakamoto_al_2009}: we address the question why the same loops appear different in different lines, which is exactly the evidence reported in \citet{Tripathi_2009}, whereas \citet{Sakamoto_al_2009} address the evidence that hot loops appear fuzzier than cooler, not co-spatial, and therefore different, loops.

There are several limitations in our approach. Small duration heat pulses are a necessary ingredient to have a multi-temperature loop system. Here we assume that the heat pulses are the same in all strands, only their timing is different, and that, at regime, they occur with a constant time average. We expect some broadness in the distribution of the heat pulses, in duration, intensity and average cadence. Other simplifications concern other assumptions on the heat pulses. We are assuming a very simple time-shape of the heat pulse, with abrupt on and off. This should not affect the results, because we analyze the strands evolution on a much longer time scale. The heat pulse is long enough not to have significative effects due to delay to reach equilibrium of ionization \citep{Reale_2008}. We also assume that each strand is heated only once, i.e. there is no reheating of the same strand. This is a realistic assumption in the framework of progressive magnetic reconnection occurring in gradually twisting loops, due to footpoint photospheric motions. Presumably there can only be one heating event per strand since it is destroyed after magnetic reconnection. Nevertheless, note that the reconnecting strands form two new (relaxed) strands which are then presumably subject to the same twisting/braiding motions which originally energised them, leading to eventual reconnection and another heating event. In principle, this process could be repeated many times for a single pair of strands. The magnetic flux in this case is not destroyed, simply reconfigured.  

The hydrodynamic description of the confined plasma should hold quite firmly in the individual loop strands, as in monolithic loops. Maybe the description of the low loop section should, for instance, include the possible tapering in the transition region (e.g. \citealt{Gabriel_1976}), but differences are expected only in the very low part of the loops and, therefore, should not affect the results presented here.

Our analysis provides also an interesting prediction: we expect high contrast to show in very hot lines, such as the Fe XXIII line and the Ca XIX line (typical flare lines), even stronger than in the cool lines. Therefore, forthcoming SDO observations and desirable future X-ray spectral observations of the quiet Sun, and in particular of active regions, at high sensitivity and enough spatial resolution, may provide an important confirmation of our scenario.

In this analysis, we have studied a coronal loop at steady-state, i.e. when its light curve and temperature is constant on average. However, our model is useful also to study the time evolution of the multistranded loops, and we will further develop our analysis in this perspective as the next step.

%
\bigskip
\acknowledgements{We acknowledge support from \emph{Italian Ministero dell'Universit\`a e Ricerca} and \emph{Agenzia Spaziale Italiana (ASI)}, contract I/015/07/0. 
 \emph{CHIANTI} is a collaborative project involving the \emph{NRL (USA)}, the \emph{Universities of Florence (Italy)} and \emph{Cambridge (UK)}, and \emph{George Mason University (USA)}.}

%
\bibliographystyle{apj}
\bibliography{biblio}

\end{document}